\documentclass[floatfix,prd,twocolumn,preprintnumbers]{revtex4}
\usepackage{graphicx}
\usepackage{amsfonts}
\usepackage{amssymb}
\usepackage{amsmath}

\newcommand*{\di}{\partial}

\begin{document}

\preprint{arXiv:0901.0892 [gr-qc]}

\title{Einstein static universes are unstable in generic $f(R)$ models}

\author{Sanjeev S.~Seahra}%
\email{sseahra@unb.ca}%
\affiliation{Department of Mathematics and Statistics, University of
New Brunswick, Fredericton, New Brunswick, E3B 5A3, Canada \mbox{ }}

\author{Christian G.~B\"ohmer}%
\email{c.boehmer@ucl.ac.uk}%
\affiliation{Department of Mathematics and Institute of Origins, 
University College London, Gower Street, London, WC1E 6BT, United Kingdom}

\begin{abstract}

We study Einstein static universes in the context of generic $f(R)$
models. It is shown that Einstein static solutions exist for a wide
variety of modified gravity models sourced by a barotropic perfect
fluid with equation of state $w=p/\rho$, but these solutions are always
unstable to either homogeneous or inhomogeneous perturbations. Our
general results are in agreement with specific models investigated
in that past.  We also discuss how our techniques can be applied to
other scenarios in $f(R)$ gravity.

\end{abstract}

\date{February 20, 2009}

\maketitle

\section{Introduction}

In 1917 Einstein tried to find a static solution of the field
equations of general relativity that describes a homogeneous and
isotropic universe. As static solutions, in general, do not exist in
this setting, Einstein introduced the cosmological constant
$\Lambda$ to make the solution static~\cite{Einstein:1917ce}. It was
noted soon thereafter by Eddington~\cite{Eddington:1930} that this
solution is unstable with respect to homogeneous and isotropic
perturbations. However, subsequent work by
Harrison~\cite{Harrison:1967zz} and Gibbons~\cite{Gibbons:1987jt}
indicates that the issue is more subtle than originally thought. In
particular Gibbons showed that the Einstein static universe
maximizes the entropy for an equation of state with sound speed $c_s
> 1/\sqrt{5}$. These results have been further investigated
in~\cite{Barrow:2003ni} where it was shown that the Einstein static
universe is neutrally stable against small inhomogeneous vector and
tensor perturbations, and also neutrally stable against
inhomogeneous adiabatic scalar density perturbations if the sound
speed satisfies $c_s > 1/\sqrt{5}$, and unstable otherwise. These
results allow for the interesting scenario where the universe may
have started out as an Einstein static
universe~\cite{Ellis:2002we,Ellis:2003qz}, thereby allowing for a
natural beginning for inflation.

Because of its interesting stability properties and its analytical
simplicity, the Einstein cosmos has always been of great interest
general relativity and its extensions.  These static cosmological
models have been constructed in braneworld models
\cite{Gergely:2001tn,Gruppuso:2004db,Seahra:2005us,Clarkson:2005mg},
string theory \cite{Hamilton:2008ey}, and loop quantum cosmology
\cite{Mulryne:2005ef,Parisi:2007kv}.  In addition, models with
non-constant pressure have been considered
\cite{Ibrahim:1976,Boehmer:2002gg,Boehmer:2003uz,Boehmer:2007ae,Lake:2008vx,Grenon:2008dg}.

The first stability analyses of the Einstein static universe in
$f(R)$ modified gravity can be found in
Refs.~\cite{Barrow:1983rx,Boehmer:2007tr}. In the latter paper, it
was found that stable solutions do exist which were unstable in
general relativity.  Subsequent work on the Einstein static universe
in higher order gravity
theories~\cite{Goswami:2008fs,Goheer:2008tn}, especially existence
and stability, had led to slightly differing results. For instance,
in~\cite{Goswami:2008fs} it was noted that that there exists only
one functional form of $f(R)$ that admits an Einstein static
universe. On the other hand, in~\cite{Boehmer:2007tr} the stability
was analyzed for a model of a different type. Both findings seem to
be inconsistent at first glance, however, in this paper we are able
to reconcile all previous results and show their consistency.  (Also
see Ref.~\cite{Clifton:2005at} for a discussion of the existence of
Einstein-static models in more general modified gravity scenarios.)

Although the cosmological constant was soon dismissed after its
original introduction in 1917, recent observations seem to indicate
that the Universe is currently undergoing a phase of accelerated
expansion, consistent with the presence of $\Lambda$ (now called
dark energy) in the Einstein field equations. The idea that dark
energy may indicate the need for a gravitational theory beyond
general relativity has recently inspired a vast amount of research,
in what is know as modified gravity or higher order gravity. Such
models are not new and have been analyzed ever since the field
equations have been formulated in their original form. However, it
is only recently that these models are investigated in the context
of alternatives to dark energy. $f(R)$ models which have a viable
cosmology were analyzed, and it was found that the models satisfying
cosmological and local gravity constraints are practically
indistinguishable from the $\Lambda$CDM model, at least at the
background level~\cite{Nojiri:2003ft,Amendola:2007nt}, for recent
reviews see~\cite{Nojiri:2008nt,Sotiriou:2008rp,Durrer:2008in}.
However, such models must also be consistent with cosmological
structure formation, which means the study of perturbation theory in
modified gravity is
necessary~\cite{Tsujikawa:2007gd,Uddin:2007gj,Bazeia:2007jj}.
Therefore, the relatively simple Einstein static universe is an
ideal test bed for perturbation theory to gain insight into the
principal differences between general relativity and its
modifications.

We analyze the stability of the Einstein static universe against
homogeneous and inhomogeneous scalar perturbations in the context of
$f(R)$ gravity. In the following section we discuss the existence of
the Einstein static universe as a solution of the cosmological field
equations for two types of models, fine-tuned and non-fined-tuned
models and make connections with previous work. In \S\ref{sec:scalar
perturbations} we analyze the perturbations about the Einstein
static universe. The perturbations can be characterized by two
parameters, namely the equation of state $w$ and a parameter
$\alpha$ which depends on the form of $f(R)$ and the matter density.
We discuss our results in the final \S\ref{disc}.

\section{Existence of Einstein static universes in $f(R)$ gravity}
\label{exis}

\subsection{$f(R)$ field equations}

In this paper, we consider modified gravity models governed by the
action
\begin{equation}
    S = \int d^4x \sqrt{-g} \left[ \frac{1}{2\kappa^2} f(R) +
    \mathcal{L}_m \right].
\end{equation}
Here, $f(R)$ is an arbitrary function of the Ricci scalar $R$,
$\mathcal{L}_m$ is the Lagrangian density of matter, and $\kappa^2 =
8\pi G$.  The field equations associated with this action are well
known:
\begin{equation}\label{field eqn}
    f'R_{ab} - \frac{1}{2} f g_{ab} + (g_{ab} \Box - \nabla_a
    \nabla_b)f' = \kappa^2 T_{ab},
\end{equation}
where $\Box = \nabla^m \nabla_m$, and we use a prime to denote
derivatives of functions with respect to their arguments; i.e., $f'
= f'(R) = df/dR$.  Note that the Ricci scalar depends on the second
derivatives of the metric, so the last term on the right-hand side of
(\ref{field eqn}) involves fourth-order derivatives of $g_{ab}$. The
trace of Eq.~(\ref{field eqn}) gives
\begin{equation}\label{trace field eqn}
    f'R -2f + 3\Box f' = \kappa^2 T.
\end{equation}
This gives a (possibly nonlinear) dynamical equation for the Ricci
scalar sourced by the trace of the stress energy tensor.  The
existence of such an equation means that instead of regarding
(\ref{field eqn}) as a fourth order equation for $g_{ab}$, we can
regard (\ref{field eqn}) and (\ref{trace field eqn}) as a system of
coupled second order equations for $g_{ab}$ and $R$.  This
observation is the basis for the treatment of $f(R)$ gravity as a
particular type of scalar-tensor theory, see
e.g.~\cite{Faraoni:2004}.

\subsection{Einstein static solutions}

To find an Einstein static solution of the field
equations~(\ref{field eqn}), we adopt the metric ansatz
\begin{equation}
    ds^2 = a_0^2 (-d\eta^2 + \gamma_{ij} d\theta^i d\theta^j),
\end{equation}
where $\gamma_{ij}$ is the metric on the 3-sphere
\begin{equation}
    \gamma_{ij}d\theta^i d\theta^j = d\chi^2 + \sin^2\chi (d\theta^2
    + \sin^2\theta \, d\phi^2).
\end{equation}
We also assume that the matter content of the model is a single
perfect fluid
\begin{equation}
    T_{ab} = (\rho+p)u_a u_b + p g_{ab}, \quad p = w\rho, \quad u^a  \di_a = \frac{1}{a_0} \di_\eta.
\end{equation}
In these, $w$ is the constant equation of state parameter and $a_0$
is the constant radius of the universe.  We have selected the
spacetime coordinates $(\eta,\chi,\theta,\varphi)$ to be
dimensionless. Putting these assumptions into the $f(R)$ field
equations yields that the Ricci scalar is equal to a constant $R_0$
fixed by the radius
\begin{equation}
    R_0 = 6 a^{-2}_0,
\end{equation}
and that the values of $f$ and its first derivative at $R = R_0$ are
\begin{equation}
    f_0 = f(R_0) = 2\kappa^2 \rho, \quad f_0' = f'(R_0) = \frac{\kappa^2
    \rho a_0^2
    (1+w)}{2}.
\end{equation}
Note that this can be rewritten as
\begin{equation}\label{ES condition}
    \left( \frac{R f'}{f} \right)_{R = R_0} = \frac{3}{2}(1+w).
\end{equation}
In other words, if we view the equation of state parameter $w$ as
fixed, then any $f(R)$ model that satisfies (\ref{ES condition}) for
one \emph{particular} value of the curvature $R = R_0$ admits an
Einstein static solution~\cite{foot1}. For example, if $M$ is some mass scale
the choice
\begin{equation}
    f(R) = M^2 \exp \left[ \frac{3(1+w)}{2} \frac{R}{M^2} \right]
\end{equation}
does not admit an Einstein static solution unless the curvature is
fine-tuned to $R = M^2$.

\subsection{GR limit with cosmological constant}

General relativity with a cosmological constant $\Lambda$ can be
recast as an $f(R)$ model with
\begin{equation}
    f(R) = R - 2\Lambda.
\end{equation}
In this case, we have an Einstein-static solution for
\begin{equation}
    \frac{R_0}{\Lambda} = \frac{6(1+w)}{1+3w}, \quad \frac{\kappa^2
    \rho}{\Lambda} = \frac{2}{1+3w}.
\end{equation}
If the equation of state parameter and cosmological constant are
given, these equations determine $R_0$ and $\rho$ uniquely; i.e., in
order to have a static configuration in general relativity the
radius of the universe must be fine-tuned.  From this it follows
that any fluctuations in the universe's radius away from this
fined-tuned value will result in time-dependent cosmologies, which
is what we will see in \S\ref{sec:scalar perturbations}.

\subsection{Models without fine-tuning}\label{sec:non-fine-tuned}

It is possible to construct $f(R)$ models such that one can find
Einstein static solutions for any choice of the radius $a_0$, or
conversely the curvature $R_0$.  To find these models, we regard
(\ref{ES condition}) as an ordinary differential equation that holds
for all curvature.  The solution is
\begin{equation}\label{eq:self-similar f(R)}
    f(R) = M^2 \left( \frac{R}{M^2} \right)^{\frac{3}{2}(1+w)},
\end{equation}
where the mass scale $M$ is an integration constant.

\citet{Goswami:2008fs} have previously considered this case and
claimed that this choice of $f(R)$ is the only one that admits
Einstein static solutions.  This statement is perhaps misleading:
(\ref{eq:self-similar f(R)}) is the only $f(R)$ model that allows
for Einstein static solutions of arbitrary radius.  In other words,
one does not need to fine-tune the radius or curvature in these
models to have an Einstein static solution, which is quite different
from the general relativity case above~\cite{foot2}. We will see in
\S\ref{sec:non-fine-tuned perts} that the spectrum of homogeneous
linear perturbations of this model admit static solutions, which is
not possible when $f(R) = R - 2\Lambda$.

\section{Perturbations}
\label{sec:scalar perturbations}

\subsection{Linearized $f(R)$}\label{sec:general EOMs}

We consider fluctuations of the background geometry and matter
content parameterized by
\begin{equation}
    \delta g_{ab} = h_{ab}, \quad \delta g^{ab} = -h^{ab},
\end{equation}
where the metric fluctuation is understood to be small $h_{ab} \ll
g_{ab}$. The variation of the Ricci tensor induced by perturbations
of the metric is
\begin{multline}\label{perturbed Ricci tensor}
    \delta R_{ac} = -\tfrac{1}{2} g^{bd} \nabla_a \nabla_c h_{bd} -
    \tfrac{1}{2} g^{bd} \nabla_b \nabla_d h_{ac} \\ + \tfrac{1}{2} g^{bd}
    \nabla_b \nabla_c h_{ad} + \tfrac{1}{2} g^{bd} \nabla_{b}
    \nabla_a h_{cd}.
\end{multline}
In all formulae, $g_{ab}$, $R_{ab}$, $f_0$, $f_0'$, etc.~refer to
background quantities, and all indices are raised and lowered with
$g_{ab}$.  From this expression, it follows that the perturbation of
the Ricci scalar is
\begin{align}\label{phi definition}
    \varphi \equiv \delta R &= -h^{ab}R_{ab} + g^{ab} \delta R_{ab}
    \nonumber\\ &= -h^{ab}R_{ab} - \Box h + \nabla_a \nabla_b h^{ab},
\end{align}
where $h = h^a{}_a$.  Note how we have defined the scalar quantity
$\varphi$ to be precisely the variation of $R$.  Finally, we have
the
linear variations in $f$ and $f'$:%
\begin{equation}
    \label{Taylor 1} \delta f = f_0' \varphi,  \quad \delta f' =
    f_0'' \varphi.
\end{equation}

We now linearize the $f(R)$ field equation (\ref{field eqn}) about
the background solution
\begin{multline}\label{graviton equation}
    f_0'' ( R_{ab} \varphi + g_{ab} \Box\varphi - \nabla_a \nabla_b \varphi) +
     f_0' (\delta R_{ab} - \tfrac{1}{2}\varphi g_{ab}) \\ - \tfrac{1}{2}
    f_0 h_{ab} + X_{ab} = \kappa^2 \delta T_{ab},
\end{multline}
where
\begin{multline}
    X_{ab} = h_{ab} \Box f_0' - g_{ab} h^{cd} \nabla_c \nabla_d f_0'
    \\ +
    (g_{ab} g^{cd} -  \delta_a{}^c\delta_b{}^d) \left\{ \left[ \tfrac{1}{2}\nabla^m h_{cd} -
    \nabla_{(c} h_{d)}{}^m \right] \nabla_m f_0' \right.
    \\ \left. + \varphi \nabla_c \nabla_d f_0'' + 2 \nabla_{(c} \varphi \nabla_{d)} f_0'' \right\}.
\end{multline}
We can also linearize the equation of motion (\ref{trace field eqn})
for $R$, which leads to
\begin{multline}\label{scalar equation}
    3f_0''(\Box - m^2)\varphi + 6\nabla_a \varphi \nabla^a f_0'' - 3h^{ab}
    \nabla_a \nabla_b f_0' \\ - 3 ( \nabla_a h^{ac} - \tfrac{1}{2}
    \nabla^c h ) \nabla_c f_0' = \kappa^2 \delta T^a{}_a,
\end{multline}
where $m$ is the effective mass of the $\varphi$ field, which is
given explicitly by
\begin{equation}\label{mass defn}
    m^2 = \frac{f_0'}{3f_0''} - \frac{R}{3} - \frac{\Box
    f_0''}{f_0''}.
\end{equation}
Equations (\ref{graviton equation}) and (\ref{scalar equation}) are
the main equations governing perturbations of generic $f(R)$ models.
Notice that they are presented as a pair of second order linear
equations for $h_{ab}$ and $\varphi$ respectively.  Using the basic
definition of $\varphi$ (\ref{phi definition}), it is possible to
combine these two equations into a single fourth order equation for
$h_{ab}$ which reflects the fourth order nature of the $f(R)$ field
equations.

\subsection{Perturbations about constant curvature
solutions}\label{sec:constant curvature EOMs}

The Einstein static solution has constant Ricci curvature, which
implies that $f_0$, $f_0'$ and $f_0''$ are independent of spacetime
position.  That is, all gradients of these quantities vanish in
equations (\ref{graviton equation}) and (\ref{scalar equation}).
These then simplify to
\begin{multline}\label{graviton equation 2}
    f_0'' ( R_{ab} \varphi + g_{ab} \Box\varphi - \nabla_a \nabla_b \varphi) +
    \\ f_0' (\delta R_{ab} - \tfrac{1}{2}\varphi g_{ab}) - \tfrac{1}{2}
    f_0 h_{ab} = \kappa^2 \delta T_{ab},
\end{multline}
and
\begin{equation}\label{scalar equation 2}
    3f_0''(\Box - m^2)\varphi = \kappa^2 \delta T^a{}_a,
\end{equation}
with
\begin{equation}\label{mass defn 2}
    m^2 = \frac{f_0'}{3f_0''} - \frac{R}{3}.
\end{equation}
Notice that these equations also govern perturbations about locally
de Sitter or anti-de Sitter solutions in arbitrary $f(R)$ models.
(Equations of motion for perturbations of de Sitter backgrounds were
first derived in \cite{Cognola:2005de}.)

\subsection{Scalar perturbations}

Working in the longitudinal gauge, we can write scalar perturbations
of the model as
\begin{subequations}\label{scalar ansatz}
\begin{eqnarray}
    h_{ab} & = & 2\Psi u_a u_b + 2\Phi (g_{ab} + u_a u_b), \\
    \delta T^{a}{}_b & = & \delta\rho u^a u_b + u^a D_b q +
    u_b D^a q + \delta p \mathcal{P}^a{}_b,
\end{eqnarray}
\end{subequations}
where $\mathcal{P}^a{}_b$ and $D_a$ are the spatial projection
tensor and derivative, respectively:
\begin{equation}
    \mathcal{P}^a{}_b = \delta^a{}_b + u^a u_b, \quad D_a =
    \mathcal{P}^b{}_{a} \di_b.
\end{equation}
In this gauge, the perturbed metric reads
\begin{equation}
    ds^2 = a_0^2 \left[ -(1-2\Psi)d\eta^2 + (1 + 2\Phi)
    \gamma_{ij} d\theta^i d\theta^j \right].
\end{equation}
From this, it follows that $\Psi$ represents the perturbation to the
Newtonian potential and $\Phi$ represents the perturbation to the
spatial curvature.  In the matter sector, $q$ is related to the
perturbation in the fluid's 4-velocity and the density and pressure
perturbations are rewritten as
\begin{equation}
    \delta\rho = \rho \delta, \quad \delta p = c_s^2 \delta\rho = w\rho \delta,
\end{equation}
where $\delta$ is the relative density perturbation and we have used
that the sound speed $c_s^2 = w$ for a single perfect fluid.

It is useful to perform a harmonic decomposition of these scalar
potentials and the perturbation to the Ricci scalar $\varphi$:
\begin{align}
    \nonumber \Psi & = \Psi_n(\eta) Y_n(\theta^i), &
    \Phi & = \Phi_n(\eta) Y_n(\theta^i), &
    \delta & = \delta_n(\eta) Y_n(\theta^i), \\
    q & = q_n(\eta) Y_n(\theta^i), &
    \varphi & = \varphi_n(\eta) Y_n(\theta^i).
\end{align}
In these expressions, summation over $n = 0,1,2,\ldots$ is
understood. The harmonic function $Y_n = Y_n(\theta^i)$ satisfies
\begin{gather}
    {}^{(3)} \Delta Y_n = -k^2 Y_n, \quad k^2 = n(n+2),
\end{gather}
where ${}^{(3)} \Delta$ is the Laplacian operator on the
3-dimensional spatial sections of the model (i.e., associated with
$\gamma_{ij}$).

Putting (\ref{scalar ansatz}) into (\ref{graviton equation 2}) and
(\ref{scalar equation 2}) and performing some algebra, we find that
$\Psi_n$, $q_n$ and $\delta_n$ are expressible in terms of $\Phi_n$
and $\varphi_n$:
\begin{gather}\nonumber
    \Psi_n = \Phi_n + \frac{f_0''}{f_0'} \varphi_n, \quad  q_n =
    \frac{f_0'' \dot\varphi_n + 2f_0' \dot\Phi_n}{\kappa^2 a_0}
    , \\ \delta_n = \frac{3f_0''\left[
    (1+w)f_0 (\ddot\varphi_n + k^2 \varphi) + 4m^2 f_0' \varphi_n \right]}{2(1-3w) f_0
    f_0'},
\end{gather}
where we have used an overdot to denote derivatives with respect to
$\eta$; i.e., $d\varphi_n/d\eta = \dot\varphi_n$.  Once these are
used to eliminate $\Psi_n$, $q_n$ and $\delta_n$, we obtain the
following equations of motion
\begin{equation}\label{scalar equations}
    \ddot{\mathbf{x}} = \mathbf{A} \mathbf{x}, \quad \mathbf{x} = \left(
    \begin{array}{c}
        \Phi_n \\ \xi_n
    \end{array}
    \right), \quad \mathbf{A} =
    \left(
    \begin{array}{cc}
        A_{11} & A_{12} \\ A_{21} & A_{22}
    \end{array}
    \right),
\end{equation}
where $\mathbf{A}$ is a constant matrix with entries
\begin{subequations}
\begin{eqnarray}
    A_{11} & = & \frac{6-k^2}{3}, \\
    A_{12} & = & \frac{\alpha (k^2+3)(1+w) +
    3}{9\mu^2(1+\alpha+w\alpha)(1+w)}, \\
    A_{21} & = & \frac{2(k^2-3)(1-3w)(1+\alpha+w\alpha)}{\alpha}, \\
    A_{22} & = & -\frac{5\alpha k^2 w + 3k^2 w^2 \alpha + 2\alpha k^2
    +6}{3(1+w)\alpha}.
\end{eqnarray}
\end{subequations}
Here we have defined
\begin{equation}
    \alpha = \frac{\kappa^2 \rho}{f_0' m^2} = \left(\frac{2{f'_0}^2}{3f_0f_o''} - 1 -w\right)^{-1}, \quad \xi_n = \frac{f_0' \varphi_n}{\kappa^2
    \rho}.
\end{equation}
The solution to the equations of motion are simple and given in
terms of four constants of integration $c_i$:
\begin{multline}
    \mathbf{x}(\eta) = \mathbf{x}_1 \left( c_1 e^{+i\omega_1 \eta} + c_2 e^{-i\omega_1 \eta}
    \right) \\ + \mathbf{x}_2 \left( c_3 e^{+i\omega_2 \eta} + c_4 e^{-i\omega_2 \eta}
    \right),
\end{multline}
where $\mathbf{x}_1$ and $\mathbf{x}_2$ are eigenvectors of
$\mathbf{A}$ corresponding to eigenvalues $-\omega_1^2$ and
$-\omega_2^2$, respectively.  Explicitly, the frequencies are given
by
\begin{equation}\label{eq:frequencies}
    \omega_1 = \sqrt{\mathcal{A} + \sqrt{\mathcal{B}}}, \quad \omega_2 = \sqrt{\mathcal{A} -
    \sqrt{\mathcal{B}}},
\end{equation}
where
\begin{subequations}
\begin{eqnarray}
    \mathcal{A} & = & \frac{1}{2} (1+w) k^2 - 1 +
    \frac{1}{(1+w)\alpha}, \\ \nonumber \mathcal{B} & =& \frac{1}{4} (1-w)^2 k^4
    + \left( \frac{1}{3} + w + \frac{1-w}{1+w} \alpha \right) k^2
    \\  & & + 6w -1 + \frac{6w}{(1+w)\alpha} + \frac{1}{(1+w)^2
    \alpha^2}.
\end{eqnarray}
\end{subequations}
Obviously, we will have unstable perturbation modes if either
$\text{Im}(\omega_1)$ or $\text{Im}(\omega_2)$ are nonzero.  This
leads to the following \emph{stability} criteria:
\begin{equation}
    \text{Im}(\omega_1) = \text{Im}(\omega_2) = 0 \,\,
    \Leftrightarrow \,\, \mathcal{A} \ge 0 \text{ and }
    \mathcal{A}^2 \ge \mathcal{B} \ge 0.
\end{equation}
In other words, if the above conditions are met for a given choice
of $(\alpha,w,k)$, the associated perturbations will be stable.  It
should be stress that the stability of a particular Einstein static
model is completely determined by $w$, $k$ and the value of $f(R)$
and its first and second derivatives \emph{evaluated on the
background solution}.  That is, it is not necessary to know the full
functional form of $f(R)$ to determine the behavior of
perturbations, we just need know about the first few terms of the
Taylor expansion of $f(R)$ about the background curvature.

\subsection{GR limit}

Notice that since $f_0'' = 0$ for $f(R) = R - 2\Lambda$, the scalar
mass $m^2 = \infty$ in the GR limit.  The corresponding limits for
the frequencies (\ref{eq:frequencies}) are
\begin{equation}\label{eq:GR frequencies}
    \lim_{\alpha \rightarrow 0} \omega_1^2 = \frac{2}{(1+w)\alpha}=\pm\infty, \quad
    \lim_{\alpha \rightarrow 0} \omega_2^2 = w(k^2 - 3) - 1.
\end{equation}
The second of these matches the results of \citet{Barrow:2003ni}.
The first frequencies are formally infinite, and represent an
artifact of the reduction of fourth-order to second-order gravity.
The second set of frequencies implies that the model is stable for
all $w(k^2 - 3) > 1$.

\subsection{Homogeneous perturbations}

In the case of homogeneous perturbations we set $k = 0$ in the above
expressions. We then find that Einstein-static solutions will be
stable in two distinct regions of parameter space. The first we call
the ``normal region'':
\begin{multline}
    \text{normal region} = \left\{ (\alpha,w) \,
    \left| \,
    \alpha \le (1+w)^{-1}(1-6w)^{-1}, \right. \right. \\ \left.
    w \le -\frac{1}{3} - \frac{1}{2\alpha} \left( 1 - \sqrt{\frac{16}{9}\alpha^2+1} \right),
    w > -1, \alpha \ge 0 \right\}.
\end{multline}
The term normal comes from the fact that the this region has $w \in
(-1,0]$, so the matter sound speed is sub-luminal.  The other
stability region is called the ``phantom region'', and is given by:
\begin{multline}
    \text{phantom region} = \left\{ (\alpha,w) \,
    \left| \, \alpha \ge (1+w)^{-1}(1-6w)^{-1}, \right. \right. \\ \left. w  < -1, \alpha < 0
    \right\}.
\end{multline}
The stable models in this region have $w < -1$ which implies a
superluminal sound speed, hence the term ``phantom''.  Also, the
fact that $\alpha < 0$ implies that either the scalar mass $m$ is
imaginary, or the density of matter $\rho$ is negative. Moreover,
the effective gravitational coupling constant is also negative.
In either possibility, models in the phantom region are quite strange
and we are led to conclude that these models should be regarded as
unphysical.

\begin{figure}
\includegraphics[width=\columnwidth]{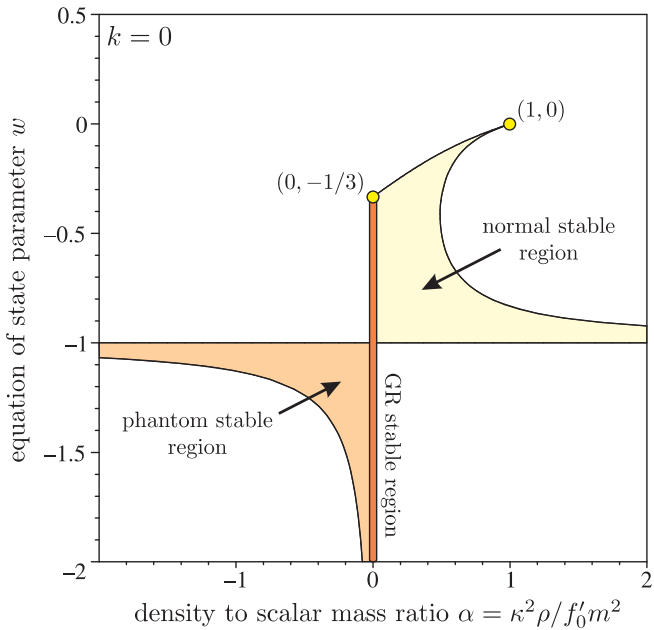}
\caption{Regions of stability in the $(\alpha,w)$ parameter space
for homogeneous perturbations of Einstein static universes. Note
that there exist unstable $k=0$ modes for all models with $w>0$.}
\label{fig:k=0}
\end{figure}
In Figure \ref{fig:k=0}, we plot the normal and phantom stability
regions for homogeneous perturbations.  We also show the general
relativity stability region as inferred from~(\ref{eq:GR
frequencies}). The key feature of this plot is that all models with
equations of state $w > 0$ are unstable with respect to homogeneous
perturbations.

\subsection{Inhomogeneous perturbations}

\begin{figure}
\includegraphics[width=\columnwidth]{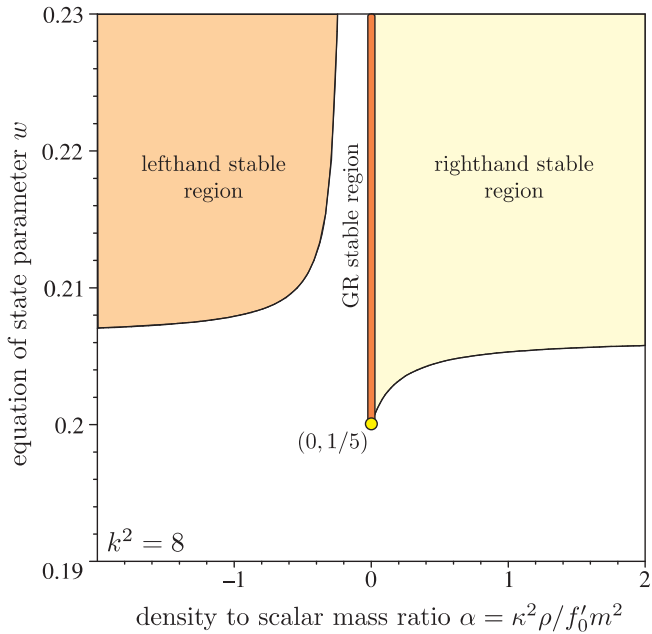}
\caption{Regions of stability in the $(\alpha,w)$ parameter space
for $k^2 = 8$ ($n=2$) perturbations of Einstein static universes.
Note that all models with $w<0$ are unstable with respect to these
types of perturbations.} \label{fig:k=sqrt(8)}
\end{figure}

We now turn our attention to the behavior of inhomogeneous
perturbations $k \ne 0$.  Recall that the spherical symmetry of the
3-dimensional spatial sections of the model dictates that the value
of the wavenumber is discrete $k^2 = n(n+2)$.  Note that as in
general relativity, the $n=1$ mode corresponds to a gauge degree of
freedom related to a global rotation.  For the $n \ge 2$ modes, we
find that the model is stable for $(\alpha,w)$ lying within two
regions:
\begin{subequations}\label{eq:inhomogeneous stability criteria}
\begin{align}
    \text{right region} & = \left\{ (\alpha,w) \, |
    \, w > \gamma_1 + \gamma_2, \alpha \ge 0 \right\}, \\
    \text{left region} & = \left\{ (\alpha,w) \, |
    \, w > \gamma_1 - \gamma_2, \alpha < 0 \right\}.
\end{align}
\end{subequations}
where
\begin{subequations}
\begin{equation}
    \gamma_1 = - \frac{3\alpha k^4 + 2(3-5\alpha)k^2 -
    6(2\alpha+3)}{6\alpha(k^4-2k^2-6)},
\end{equation}
and
\begin{multline}
\gamma_2^2 = (6\alpha)^{-2} \left( {k}^{4}-2 {k}^{2} -6 \right)
^{-2} [ 9 {k}^{8}{\alpha}^{2}-
\\ 12\alpha \left( \alpha-3 \right) {k}^{6} - 4 \left( 35{\alpha}^{2}+ 39 \alpha- 9
\right) {k} ^{4}+ \\ 24 \left( 4\alpha^2+ 3\alpha-9 \right) {k}^{2}
+324+576  {\alpha}^{2} ].
\end{multline}
\end{subequations}
In Figure \ref{fig:k=sqrt(8)} we show the lefthand and righthand
stability regions for the largest wavelength (non-gauge)
inhomogeneous perturbations $k^2 = 8$.  As in Figure \ref{fig:k=0},
we also show the relevant stability condition for general
relativity.

From Figure \ref{fig:k=sqrt(8)}, we see that all models with $w < 0$
are unstable to $n =2$ perturbations.  It is not hard to confirm
that this it true for all $n \ge 2$ perturbations.  It is clear that
the stability regions in Figures \ref{fig:k=0} and
\ref{fig:k=sqrt(8)} do not overlap.  From this we can conclude that
it is impossible to construct an Einstein static universe in $f(R)$
gravity that is stable with respect to both homogeneous and
inhomogeneous perturbations, which is the main conclusion of this
paper.

\subsection{Non-fine-tuned models}\label{sec:non-fine-tuned perts}

As an example of the general results we have derived, we examine the
non-fine-tuned models of \S\ref{sec:non-fine-tuned}.  Using the
specific form of $f(R)$ given in (\ref{eq:self-similar f(R)}), we
can easily obtain $\alpha$ as a function of $w$ for these models:
\begin{equation}\label{eq:alpha non-fine-tuned}
    \alpha = \frac{1+3w}{(1+w)(1-3w)}.
\end{equation}
For the homogeneous ($k=0$) perturbations, this leads to the
following frequencies:
\begin{equation}
    \omega^2_{1,2} = -\frac{6w}{1+3w} \left[ 1 \pm
    \text{sgn}\left( w^2 - \frac{1}{3} w \right) \right].
\end{equation}
It is clear that for any choice of $w$, one of these frequencies is
zero and the other is nonzero.  This means we can always find static
and homogeneous perturbations of these models.  Recall that the
unique feature of the non-fine-tuned scenario is that there exists
Einstein static solutions for all radii $a_0$.  Hence, these static
solutions are easy to understand: they represent the deformation of
one Einstein static solution into another one with a different
radius. We can also see that the non-zero frequency will be real
only for $w \in (-1/3,0]$; i.e., the $k=0$ perturbations will be
stable when $w \in (-1/3,0]$.

When the particular form of $\alpha$ in (\ref{eq:alpha
non-fine-tuned}) is inserted into the inhomogeneous stability
criteria (\ref{eq:inhomogeneous stability criteria}), we find that
all $n \ge 2$ perturbations will be stable if
\begin{equation}
    w \ge \frac{\sqrt{5}-1}{6}.
\end{equation}
This reproduces the result of~\citet{Goswami:2008fs}, and hence
provides a nice verification of our general formulae.

\section{Conclusions}
\label{disc}

In this paper we showed that one cannot construct an $f(R)$ modified
gravity theory such that an Einstein static universe is stable with
respect to both homogeneous and inhomogeneous scalar perturbations.
Our results and general formulae are valid for any generic form of
the function $f(R)$. Our approach allows us to verify results
previously derived based on specific forms of $f(R)$,
see~\cite{Boehmer:2007tr}. Furthermore, we were able to reconcile
the results of~\cite{Boehmer:2007tr} and~\cite{Goswami:2008fs},
which at first glance seem to be contradictory.

Our results show explicitly that perturbation theory of modified
gravity theories shows a much richer stability/ instability
structure than general relativity. This due to the fundamentally
fourth order nature of the theory, as evidenced by the $2\times 2$
matrix equation of motion (\ref{scalar equations}) for scalar
perturbations.  This should be compared to the equation of motion in
general relativity (as in \cite{Barrow:2003ni}, for example), which
is just a single second-order ODE.  In case of homogeneous
perturbation we found a region in the parameter space where the
equation of state is in the phantom regime $w < -1$, see the phantom
stable region of Fig.~\ref{fig:k=0}. It is unclear how this
stability region can be interpreted from a physical point of view.

Finally, we would like to point out how the formulae and techniques
in this paper can be applied to other situations.  We note that the
equations of \S\ref{sec:general EOMs} can be applied to any $f(R)$
background solution.  The formulae of \S\ref{sec:constant curvature
EOMs} can be used for any constant curvature solution of the
modified gravity field equations; i.e., they could be used to
address the stability of black hole solutions of the modified field
equations.  Our method of determining stability criteria entirely in
terms of the background values of $f$ and its first and second
derivatives is applicable beyond the Einstein static solution.
Indeed, it can be an efficient method of analyzing perturbations of
given constant curvature spacetime in the entire space of $f(R)$
models.

\acknowledgments We thank Peter Dunsby and Naureen Goheer for useful
discussions. SSS is supported by NSERC.

\end{document}